\documentclass[aps, prl, preprintnumbers,notitlepage, superscriptaddress, twocolumn]{revtex4-1}
\usepackage{amsmath, amsthm, amssymb}
\usepackage{mathpazo}
\usepackage{hyperref}
\hypersetup{
    colorlinks,
    citecolor=magenta,
    filecolor=black,
    linkcolor=blue,
    urlcolor=black
}
\newcommand{\be}{\begin{equation}}
\newcommand{\ee}{\end{equation}}
\newcommand{\ben}{\begin{enumerate}}
\newcommand{\een}{\end{enumerate}}
\def \bea{\begin{align}}
\def \eea{\end{align}}

\newcommand{\half}{\frac{1}{2}}

\newcommand{\ID}{{\mathbb{1}}}

\newcommand{\ra}{\rangle}
\newcommand{\la}{\langle}
\newcommand{\da}{\downarrow}
\newcommand{\ua}{\uparrow}

\makeatletter

\newcommand{\Rmnum}[1]{\expandafter\@slowromancap\romannumeral #1@}
\makeatother

\newtheorem{thm}{Theorem}
\newtheorem{lemma}{Lemma}

\newtheorem{obs}[thm]{Observation}
\fontfamily{palatino-ttf}

\def\li{\left}
\def\pr{\right}

\providecommand{\psimfull}[2]{{\cal P}_{sim}\li(#1,#2\pr)}
\providecommand{\psimf}{{\cal P}_{sim}}

\providecommand{\pstd}[2]{{\cal P}\li(#1,#2\pr)}

\providecommand{\ppar}[2]{{\cal P}^{\circ #1}_{par}(#2)}
\providecommand{\ppark}{{\cal P}^{\circ k}_{par}}

\providecommand{\precc}{{\cal P}_{rec}}

\def\tr{{\rm Tr}}
\def\id{{\operatorname{id}}}
\def\ra{\rangle}
\def\la{\langle}
\def\>{\rangle}
\def\<{\langle}

\def\hs{{\cal H}}

\def\squareforqed{\hbox{\rlap{$\sqcap$}$\sqcup$}}
\def\qed{\ifmmode\squareforqed\else{\unskip\nobreak\hfil
\penalty50\hskip1em\null\nobreak\hfil\squareforqed
\parfillskip=0pt\finalhyphendemerits=0\endgraf}\fi}

\providecommand{\norm}[1]{\lVert#1\rVert}

\begin{document}

\title{Generalized teleportation and entanglement recycling}

\author{Sergii Strelchuk}
 \email{ss870@cam.ac.uk}
\affiliation{Department of Applied Mathematics and Theoretical Physics, University of Cambridge, Cambridge CB3 0WA, U.K.}
\author{Micha\l{} Horodecki}
\affiliation{Institute for Theoretical Physics and Astrophysics, University of Gda\'nsk, 80-952 Gda\'nsk, Poland}
\author{Jonathan Oppenheim}
\affiliation{Department of Applied Mathematics and Theoretical Physics, University of Cambridge, Cambridge CB3 0WA, U.K. \\ University College of London, Department of Physics \& Astronomy, London, WC1E 6BT and London Interdisciplinary Network for Quantum Science}

\begin{abstract}
We introduce new teleportation protocols which are generalizations of the original teleportation protocols that use the Pauli group~\cite{bennett_teleporting_1993} and the port-based teleportation protocols, introduced by Hiroshima and Ishizaka~\cite{ishizaka_asymptotic_2008}, that use the symmetric permutation group. We derive sufficient condition for a set of operations, which in general need not form a group, to give rise to a teleportation protocol and provide examples of such schemes. This generalization leads to protocols with novel properties and is needed to push forward new schemes of computation based on them.
%opens up the perspective for new methods of computation. 
Port-based teleportation protocols and our generalizations use a large resource state consisting of $N$ singlets to teleport only a single qubit state reliably. We provide two distinct protocols which recycle the resource state to teleport multiple states with error linearly increasing with their number. The first protocol consists of sequentially teleporting qubit states, and the second teleports them in a bulk.

%The first teleportation protocol that does not require correction after the measurement is port-based teleportation introduced by Hiroshima and Ishizaka. It uses a large resource state consisting of $N$ singlets to teleport only a single qubit state reliably. We provide two distinct protocols which recycle the resource state to teleport multiple states with error linearly increasing with their number. The first protocol consists of sequentially teleporting qubit states, and the second one teleports them in a bulk. 
%
%Next, we discuss possible generalizations of existing teleportation protocols. Thus far, all known protocols can be viewed as teleportation over the Pauli group or the symmetric permutation group. We derive sufficient condition for a set of operations, which in general need not form a group to be used in port-based teleportation protocols, potentially leading to new teleportation protocols with novel properties. This opens up the perspective for new methods of computation.

\end{abstract}
\maketitle

Teleportation lies at the very heart of quantum information theory, being the pivotal primitive in a variety of tasks. Teleportation protocols are a way of sending an unknown quantum state from one party to another using a resource in the form of an entangled state shared between two parties, Alice and Bob, in advance. First, Alice performs a measurement on the state she wants to teleport and her part of the resource state, then she communicates the classical information to Bob. He applies the unitary operation conditioned on that information to obtain the teleported state.

A notable use of teleportation is in relation to computing, where it plays a key role enabling universal quantum computation and establishing a strong link between a particular teleportation protocol and a kind of computation possible to be implemented using it~\cite{gottesman_demonstrating_1999}.

Recently, Hiroshima and Ishizaka introduced {\it port-based teleportation}~\cite{ishizaka_asymptotic_2008} which has the distinct property that Bob does not need to apply a correction after Alice's measurement. It is an important primitive for programmable quantum processors~\cite{ishizaka_asymptotic_2008, ishizaka_quantum_2009,nielsen_programmable_1997,brukner_probabilistic_2003}, which rely on an efficient way of storing a unitary transformation and acting it on an arbitrary quantum state.  This protocol evades the fundamental limitations of the no-go theorem proved in~\cite{nielsen_programmable_1997}, which states that universal deterministic programmable quantum processors cannot exist. Even though the protocol makes it possible to execute arbitrary instructions deterministically, the result will be inherently noisy.

Port-based teleportation has already found its use in instantaneous non-local quantum computation~\cite{bk2011}. In the latter task, using it as the underlying teleportation routine dramatically reduced the amount of entanglement required to perform it. Such computations proved to be instrumental in attack schemes on position-based quantum cryptography~\cite{kent_quantum_2010, buhrman_garden-hose_2011, buhrman_position-based_2011, brassard_quantum_2011}. Currently, it is known that the minimum amount of entanglement an adversary needs to perform a successful attack on the scheme must be at least linear in the number of communicated qubits~\cite{buhrman_garden-hose_2011}. Also, an adversary having access to at most an exponential amount of entanglement can successfully break any position-based cryptography scheme~\cite{bk2011}. However, we do not know how much entanglement is necessary to break all schemes of this kind. Any improvement of the underlying teleportation protocol will invariably lead to the decrease of amount of entanglement required to break them, and potentially render such attacks more feasible. 

Port-based teleportation works as follows: at the beginning of the protocol Alice and Bob share a resource state, which consists of $N$ singlets $|\Psi^-\ra_{AB}=\frac{1}{\sqrt{2}}(|01\ra-|10\ra)$, termed {\it ports}. Alice performs a measurement in the form of POVM on the joint system, that includes the state she wants to teleport and her resource state. She obtains the measurement outcome $i$ from $1$ to $N$ and communicates it to Bob, who traces out all the port subsystems except for $i$-th one, discarding the remaining entanglement. The $i$-th port now contains the teleported state. 

 Although conceptually appealing, port-based teleportation relies on the properties of the symmetric permutation group, which limits its scope. In particular, it restricts the use of such teleportation protocols to implement gates specific to the underlying group. In the case of ordinary teleportation these gates correspond to Clifford-type computation~\cite{gottesman_demonstrating_1999}.
Another drawback of port-based teleportation is that it requires an enormous amount of entanglement in the resource state to teleport a single quantum state with high fidelity. This makes it extremely ill-suited for practical purposes. Decreasing the amount of entanglement required to teleport a sequence of quantum states will result in more efficient storage of the program encoded in unitary transformation as well as making efficient instantaneous non-local quantum computation, and tasks that depend on it.  

In this Letter we address the two issues above. First, we find sufficient condition for the generalized teleportation protocols, which is needed to push forward new schemes of computation based on them. Second, we introduce a recycling scheme, which drastically reduces the amount of entanglement used in port-based teleportation, and therefore allows for efficient attacks on position based cryptography. 

To tackle the first problem, we find sufficient condition that Alice's operations have to satisfy in order to make them amenable to be used in more general teleportation protocols and provide examples. From a group-theoretic perspective all currently known teleportation protocols can be classified into two kinds: those that exploit the Pauli group~\cite{bennett_teleporting_1993} and those, which use the symmetric permutation group~\cite{ishizaka_asymptotic_2008}. Such a simple change of the underlying group structure leads to two protocols with striking differences in the properties: the former protocol uses a finite resource state to teleport the state perfectly, but the receiver must make the correction to obtain the state, whereas the latter protocol require an infinitely big resource state to teleport the state perfectly, while not needing a correction on the receiver's side. The former teleportation scheme was used in the celebrated result of Gottesman and Chuang~\cite{gottesman_demonstrating_1999} to perform universal Clifford-based computation using teleportation over the Pauli group. The generalized teleportation protocol introduced in this Letter embraces both known protocols, and paves the way for protocols which lead to programmable processors capable of executing new kinds of computation beyond Clifford-type operations. The operations in the generalized teleportation protocol need not form a group. Also, because teleportation is known to be intimately connected to the variety of other fundamental tasks in quantum information processing~\cite{werner_all_2001}, its generalized version brings the potential for protocols with new properties, which depend on its implementation. 

To address the second problem we introduce two distinct protocols, which recycle the entanglement available in the resource state. Using a single resource state comprised of $N$ ports, they teleport any number of systems which is sublinear in $N$ with an error that linearly increases with the number of teleported states. The first protocol amounts to sequentially teleporting qubit states, recycling the original resource state. This can be viewed as the application of the original port-based teleportation with the resource state, followed by a resource recycling step. The resource degrades with every teleported state. In the second protocol Alice teleports her states in one go, performing the POVM, which randomly assigns each of the teleported states to one of the ports. The latter protocol, rather remarkably, provides the same finite case and asymptotic performance as the former: both of the protocols operate with an error, which is linear in the number of systems teleported. A similar idea about recycling the entangled state was used in the context of a remote state preparation protocol~\cite{bennett_remote_2005}.
The ability to recycle entanglement in such protocols has an immediate effect on the entanglement consumption of the instantaneous computation and position-based cryptography: an adversary may conduct an attack on any position-based cryptography scheme using a linear amount of entanglement in the number of communicated qubits for the case when communicating parties are constrained to product measurements. 

{\bf Generalized Teleportation.}
Until now, group-theoretic aspects of the teleportation protocols were largely overlooked. Currently, there are two distinct groups, which undergird different teleportation protocols. The first one is the Pauli group, which appeared in the first teleportation protocol of Bennett et al.~\cite{bennett_teleporting_1993}. Another one, the symmetric permutation group ${\cal S}_N$ was implicitly used in the port-based teleportation protocol of~\cite{ishizaka_quantum_2009, ishizaka_asymptotic_2008}. Therefore, we recast the description of the port-based teleportation protocol to elicit its connection with ${\cal S}_N$, and provide the basis for generalized teleportation protocols.

This port-based teleportation protocol~\cite{ishizaka_asymptotic_2008} can be equivalently viewed as such where Alice applies a measurement, which corresponds to the action of some element $g$ from some set $G$ on her total state. In the next step, Alice sends the description of $g$ to Bob who then applies the unitary transformation $U_g^\dagger$ conditioned on $g$ to his overall state, to reach some predefined terminating state. 
We say that the teleportation protocol $\cal P$ successfully {\it terminates} when Bob obtains the state $\sigma_B\otimes\phi_{B_0}$, where $\phi_{B_0}$ is the teleported state, and $\sigma_B$ is the state of the remaining ports. In the case of port-based teleportation $U_g$ acts as a swap operation between the port where the state was teleported and the first port.

Now we consider the generalized form of the teleportation protocol where all the operations on Alice are members of some set $G$, $|G|=K$, which in general need not form a group. The protocol that is able to teleport an unknown quantum state reliably under Alice's operations which belong to the set $G$ is denoted as ${\cal P}^{G}$. Recall that the task of teleportation is in correspondence with the problem of signal discrimination for qudits~\cite{bk2011}: the probability $p_s(G)$ of successfully discriminating a set of signals $\{\eta_g\}_{g\in G}$, where
\be\label{groupsignals}
\eta_g = U_g\li(\tr_{B_1...B_N\backslash B_{g}}|\Psi_{in}\ra\la\Psi_{in}|_{AB}\pr)U_g^\dagger
\ee
after Alice applied her operation is related to the fidelity of teleportation protocols in the qudit case as $F({\cal P})=\frac{K}{d^2}p_s(G)$.

In the generalized protocol, parties start with the resource state $|\Psi_{in}\ra_{AB}=\otimes_{i=1}^{N}|\Psi^-\ra_{A_iB_i}$, and perform the following steps:
\begin{enumerate}\setlength{\itemsep}{.1pt}
\item Alice applies $\Pi_g\otimes \ID_B\li(\phi_{A_0}\otimes (\Psi_{in})_{AB}\pr)=\theta_{A_0AB}$, where $\Pi_g = |\eta_g\ra\la\eta_g|$, ${g\in G}$.
\item Alice communicates the identity of the element $g$ to Bob.
\item Bob applies $U_g^{\dagger}$ to his subsystems.
\end{enumerate}

The following Lemma presents the sufficient condition which Alice's operations must satisfy in order to induce the reliable teleportation scheme:

\begin{lemma}\label{groupteleplemma}
Define $\eta_{avg}=\frac{1}{K}\sum_{g\in G}\eta_g$. For all $G$, the protocol ${\cal P}^G$ reaches terminal state $\Omega_B$ such that $\norm{\Omega_B-\sigma_B\otimes\phi_{B_0}}_1\le\epsilon$ with $\epsilon\to0$ in the limit $N\to\infty$ if 
\be\label{condition}
\tr\li[\eta_{avg}\pr]^2\le\frac{1}{(1-\epsilon){d^{N+1}}},
\ee
where $d$ denotes the dimension of each of the subsystem.
\end{lemma}
The proof of Lemma~\ref{groupteleplemma} is located in Section 1 of the Supplemental Material.

A particular example of the unitaries, which possess the property required by Lemma~\ref{groupteleplemma} is any 2-design~\cite{dankert_exact_2009} $\{U_g\otimes U_g\}_{g\in G}$ based on some group $G$. Another example of the set $\{U_g\}_{g\in G}$ that induces $\{\eta_g\}_{g\in G}$ is the set of random unitaries introduced in~\cite{randomizingstates}, and it is easy to construct plenty of others.
%The unitaries induced by the set $G$, which satisfies~\eqref{condition} in the Lemma~\ref{groupteleplemma} must comprise a 2-design.

A particular example of the set $\{U_g\}_{g\in G}$ that induces $\{\eta_g\}_{g\in G}$ is the set of random unitaries introduced in~\cite{randomizingstates}, and it is easy to construct plenty of others.

{\bf{Recycling of the resource state}}. We now introduce two schemes that recycle entanglement in the resource state. Our first protocol consists of sequentially teleporting a sequence of qubits using a pre-shared resource state, which is made of $N$ singlets. One can view it as the multiple application of the port-based teleportation protocol introduced in~\cite{ishizaka_asymptotic_2008}, where instead of getting rid of the resource state in the end of the protocol, the parties keep it. For the programmable processor, this corresponds to executing instructions using a simple queue. To ensure that the protocol is indeed capable of teleporting multiple states while recycling the original resource state, it suffices to show that the latter does not degrade much. We do so by finding that the upper bound on the amount of distortion the resource state incurs after the next teleportation round is small, or, equivalently, we find that the fidelity of the resource state with the maximally entangled state does not change much with recycling. More formally, consider Alice and Bob who start with the initial state $|\rho_{port}\ra =\otimes_{i=1}^{N}|\Psi^-\ra_{A_iB_i}$. We will henceforth refer to each $A_iB_i$ as a ${\it port}$, with the subsystems $A_i$, $B_i$ being held by Alice and Bob respectively. 
In addition, they hold a state $\rho_{A_0R_0}=|\Psi^{-}_{A_0R_0}\ra\la\Psi^{-}_{A_0R_0}|$, and Alice wants to teleport the state of subsystem $A_0$ to Bob with $R_0$ serving as a reference system which neither party has access to. 
The total state (resource state together with the state to be teleported) they share at the beginning of the protocol is $|\Psi_{in}\ra = |\Psi^-_{A_0R_0}\ra\otimes|\rho_{port}\ra$.

We define the recycling protocol $\precc$ to be the following sequence of actions:
\begin{enumerate}\setlength{\itemsep}{.1pt}
\item Alice performs a measurement $\Pi_i$ with $\sum_{i=1}^{N}\Pi_i=\mathbb{1}_{A_0...A_N}$, getting an outcome $z=1...N$. Port $z$ now contains the teleported state. 
\item Alice communicates $z$ to Bob.
\item Bob applies a SWAP operator to ports $z$ and 1.
\item Alice and Bob mark port $1$ and do not use it in the next rounds of teleportation.
\item Alice and Bob repeat steps 1-4 using unmarked ports.
\end{enumerate}
As in the original deterministic teleportation protocol from~\cite{ishizaka_asymptotic_2008}, in step 1 Alice performs a measurement (POVM) on $A_0...A_N$ with elements $\Pi_i = \rho^{-\half}\sigma^{(i)}\rho^{-\half}$, where  $\rho = \sum_{i=1}^{N}\sigma^{(i)}$, and $\sigma^{(i)} =\frac{1}{2^{N-1}}P^-_{A_0A_i}\otimes\mathbb{1}_{\overline{A_0A_i}}$; $P^-_{A_0A_i}$ is the projector onto $\Psi^-_{A_0A_i}$. We will further adopt the notation $\overline{A_i}=A_1...A_N\backslash A_i$. To determine the success of the subsequent rounds of teleportation we compare the state of all of the ports $|\Psi_{out}^i\ra$ after Alice measures $\Pi_i$ with the special reference state $|\Psi_{id}^i\ra= |\Psi^-_{A_0A_i}\ra|\Psi^-_{R_0B_i}\ra\otimes_{j=1,j\neq i}^{N}|\Psi^-_{A_jB_j}\ra$, which corresponds to the idealized situation when the successful teleportation is carried out without any disturbance to the remaining ports. 

To show that after the first three steps of $\precc$ the state of the remaining ports is sufficiently good to be recycled in further teleportation rounds, it is enough to demonstrate that the output state $\rho_{out}^i=|\Psi_{out}^i\ra\la\Psi_{out}^i|$ has high average fidelity with $\Psi^i_{id}=|\Psi_{id}^i\ra\la\Psi_{id}^i|$:
\be
F\li(\precc\pr) = \sum_{i=1}^{N}p_iF\li(\rho^i_{out}, \Psi^i_{id}\pr),
\ee
where the superscripts in $\rho_{out}^i,\Psi^i_{id}$ denote the corresponding states after the teleported state goes to port $i$, and the last term denotes the probability that the teleportation fails.

Our first result is that the protocol $\precc$ does not degrade the total resource state by much:
\begin{thm}\label{lemma:composability}
After the steps 1-4 of $\precc$:
\be
F\li(\precc\pr)\ge 1-\frac{11}{4N} + O\li(\frac{1}{N^2}\pr).
\ee
\end{thm}
The proof of the Theorem is located in Section 2 of the Supplemental Material. We will further omit the quadratic terms in the bounds. 

Once we have established that it is possible to recycle the resource state, it is important to understand how the error accumulates after each round of teleportation. When the number of ports $N$ and rounds $k$ is relevant we denote it together with the protocol as $\precc(N,k)$. It turns out that Alice and Bob can guarantee that the error is at most additive in the number of rounds:
\begin{lemma}\label{lemma:errorlemma}
After teleporting $k$ qubits the resulting fidelity is lower bounded as:
\be
F(\precc(N,k))\ge 1-\frac{11k}{2N}.
\ee
\end{lemma}
The proof of Lemma~\ref{lemma:errorlemma} is located in Section 2 of the Supplemental Materials.

{\bf Simultaneous teleportation.} We now present the second protocol, which recycles the entanglement in the resource state much differently to that of the first one. 
Consider Alice, wishing to teleport $k$ qubits simultaneously to Bob. Parties share the resource state $|\rho_{port}\ra=\otimes_{i=1}^{N}|\Psi^-\ra_{A_iB_i}$, and Alice wants to teleport the systems $A_0...A_k$. The protocol for simultaneous teleportation is similar to steps 1-3 of $\precc$, with the following changes. Instead of $N$ POVM elements, there are $\frac{N!}{(N-k)!}$ of them, each corresponding to the possible ports that the teleported states could appear in. After the measurement, instead of a single port number Alice reveals the identity of $k$ ports where the $k$ states went to. We denote the protocol that uses $N$ ports and teleports $k$ qubits simultaneously as $\psimfull N k$.

Theorem~\ref{lemma:ksimlemma} shows that this protocol can indeed teleport $k>1$ states at once efficiently. From the Theorem it follows that the resource state degrades proportionally to the number of qubits teleported.  
\begin{thm}\label{lemma:ksimlemma}
The fidelity of simultaneous teleportation of $k$ qubits using steps 1-5 of the port-based teleportation protocol above is
\be
F(\psimfull N k)\ge 1-\frac{4k}{N}.
\ee
\end{thm}
The proof of the Theorem is located in the Section 3 of the Supplemental Material.

One can see that in the limit $N\to\infty$ the teleportation scheme works with perfect fidelity when the number of systems that Alice can teleport is sublinear in $N$.

{\bf Parallel repetition of port-based protocol}. 
In addition to the two protocols above, we introduce the protocol, which makes it possible for concurrent teleportation of the states from Alice to Bob which does not require recycling of the original state. It does so by means of partitioning the resource state into smaller parts and running the original port-based teleportation~\cite{ishizaka_asymptotic_2008} on each of the parts independently. More precisely, the protocol, denoted as $\ppar{k}{N}$, consists of teleporting $k$ qubits by running port-based teleportation protocol $k$ times in parallel each utilizing $\frac{N}{k}$ ports each time to teleport a single qubit. We will see that this protocol is substantially worse than the previous two.

{\bf Performance of the port-based protocols}.
Let us now bring together $\precc(N,k)$, $\psimf(N, k)$ and $\ppar{k}{N}$, in order to compare their performance in the task of teleporting $k$ states when the resource state consists of $N$ ports. To show how they stack up against each other we introduce a common measure of the performance of the protocols in the following definitions:

\begin{defin}the port-based teleportation protocol $\pstd{N}{q}$ is said to be {\it reliable} if it requires $N$ ports (singlets) to teleport a sequence $q\equiv q(N)$ of qubits with fidelity of teleportation satisfying
\be
\lim_{N\to\infty}F(\pstd{N}{q})=1.
\ee\end{defin}

\begin{defin}we say that the reliable protocol $\pstd N q$ is {\it efficient} if it can teleport $Q_{\cal P}(N)=\operatorname*{arg\,max}_{q} \pstd N q$. 

One can establish a partial order on the set of efficient protocols: the protocol ${\cal A}=\pstd {N} {q_1}$ is more efficient than ${\cal B}=\pstd N {q_2} $ (denoted as $ {\cal A}(N,q_1)\ge {\cal B}(N,q_2)$) if there exist sequences $Q_{\cal A}(N), Q_{\cal B}(N)$ such that $\exists N_0 \forall N\ge N_0:$ $Q_{\cal A}(N)\ge Q_{\cal B}(N)$.
\end{defin}

{\it The Lower bounds.}

The achievable fidelity of the total teleportation of $\ppar{k}{N}$ is
\be
F(\ppar{k}{N})=\li(1-\frac{3k}{4N}\pr)^{k}\ge 1-\frac{3k^2}{4N}.
\ee
Therefore, the lower bound for the performance of the protocol is:
\be
{\cal Q}_{\ppark}(N) \ge o\li(\sqrt{N}\pr).
\ee

From Lemma~\ref{lemma:errorlemma} it follows that by using $\precc$ we can teleport at least a sublinear number of qubits in the number of ports reliably, thus:
\be
{\cal Q}_{\precc}(N)\ge g(N),
\ee
where $g(N)\in o(N)$. Lastly, for $\psimfull N k$ using the result of Theorem~\ref{lemma:ksimlemma} we get:
\be
{\cal Q}_{\psimf}(N)\ge g(N),
\ee
where $g(N)\in o(N)$. Even though $\precc(N,k)$ and $\psimfull N k$ are the protocols with completely dissimilar modes of operation, and being, strictly speaking, incomparable, they achieve the same asymptotic figure of merit -- teleporting a sublinear number of systems. While both protocols achieve perfect fidelity of teleportation in the limit, one cannot be reduced to another, as they use the resource state for teleportation in an entirely different way. In the former protocol Alice applies a POVM that induces a permutation, which assigns the teleported state to one of the ports and communicates its identity to Bob via the classical channel. In the latter one, she applies a single 'large' permutation that assigns each of the $k$ teleported qubits to some unique port, followed by a single round of classical communication. The action of the permutation in $\psimfull N k$ cannot always be simulated by the repeated application of the permutation and classical communication from $\precc (N,k)$, because permutations do not commute in general.

{\it Upper bound.}
The way we approached the calculation of the fidelity of teleportation in all of the protocols enabled us to find lower bounds  for each of the protocols, but it gave no insight as to whether they are optimal. In what follows, we present a simple protocol-independent upper bound based on no-signalling principle.

\begin{obs}\label{upperb}
For any port-based teleportation protocol $\pstd N k$ we have
\be
{\cal Q}_{\cal P}(N) \le \frac{N}{2}.
\ee
\end{obs}
To justify this bound, consider a generalized port-based teleportation protocol where at the beginning Alice randomly picks one of two states to teleport: $|\Psi_0\ra = |0\ra^{\otimes k}$ or $|\Psi_1\ra = |1\ra^{\otimes k}$. She performs a measurement prescribed by the protocol and is yet to communicate its outcome to Bob. If the protocol succeeds in transmitting $k>\frac{N}{2}$ reliably, then there is no need to send any classical communication to Bob because he could measure each of the ports getting outcomes $0$ and $1$, and taking the majority vote to determine the teleported message with certainly. However, this is impossible, as it violates the no-signalling principles, which prohibits superluminal communication between Alice and Bob. Therefore, the maximum number of qubits that Alice can reliably communicate to Bob using port-based protocol is 
\be
k\le\frac{N}{2}.\label{maxrate}
\ee

It is an intriguing open question -- which structures satisfy the sufficient condition of the Lemma~\ref{groupteleplemma}, and, more importantly, what novel forms of computation might lead from here. In other words, what sets of unitaries $\{U_g\}_{g\in G}$ lead to interesting computation schemes.
An important open question is whether
one can find a set of such unitaries which allow for new teleportation
based computation schemes beyond those considered in~\cite{gottesman_demonstrating_1999}.

Finally, having established the possibility of recycling and simultaneous teleportation in the port-based protocols, makes the implementation of the programmable processors more feasible, as one can now carry out the operations using less entanglement. The true potential of these protocols is yet to be fully explored.

{\it Acknowledgements.}
We thank Fernando Brand\~ao for suggesting the protocol of simultaneous teleportation $\psimf$. Along with Matthias Christandl, he independently performed a similar calculation for its lower bound. 
S.S. thanks Trinity College, Cambridge, for its support throughout his Ph.D. studies.  M.H. is supported by EU grant QESSENCE, the project QUASAR of the National Centre for Research and Development of Poland and by Polish Ministry of Science and Higher Education grant N202231937. J.O. acknowledges the support of the Royal Society.

\vspace{2cm}
{\centerline{\bf{\large Supplemental Material}}}
\section{1. Generalized Teleportation}
{\bf Proof of Lemma 1}:

Consider the expression for teleportation fidelity expressed in terms of hypotheses testing in Lemma A.1 of~\cite{bk2011}:
\be\label{AA}
F({\cal P})=\frac{K}{d^2}p_s(G),
\ee

where $p_s(G)$ is the probability to successfully distinguish a set of states induced by Alice's operations from $G$ (the case when $K=N$ and $G=S_N$ was originally introduced in~\cite{ishizaka_asymptotic_2008}). The expression for $p_s(G)$ can be conveniently presented in terms of the states, induced by operations from $G$:
\begin{align}
p_s(G)&\ge\frac{1}{K}\li[\frac{1}{\frac{1}{K} (\sum_i\mbox{rank}\eta_g)}\frac{1}{\tr(\eta_{avg})^2}\pr]\nonumber\\
&=\frac{1}{K}\frac{1}{d^{N-1}}\frac{1}{\tr(\eta_{avg})^2}.\label{BB}
\end{align}
where $\eta_g$ have the form of (12) in the main body.

If operations in $G$ that achieve the terminal state $\Omega_B$ with $\norm{\Omega_B-\sigma_B\otimes\phi_{B_0}}_1\le\epsilon$, then at the end of teleportation protocol the fidelity is at least: 
\be
F({\cal P})\ge1-\epsilon.
\ee

Substituting expressions from~\eqref{AA} and~\eqref{BB} we see that the action of the set $G$ on the ports must be
\begin{align}
\tr(\eta_{avg})^2\le (1-\epsilon)d^{N+1}
\end{align}
to achieve the reliable teleportation. This proves the result.
\qed

\section{2. Entanglement Recycling}
We first find the expression for the total state after Alice's measurement and the ideal state respectively. It is followed by the explicit expression of the fidelity, which precedes the proof of Theorem 1. For convenience of notation, we will further label the reference system $R_0$ to be $B_0$.

The application of $\Pi_i$ results in 
\be\label{afterstate}
|\Psi_{out}^i\ra = \frac{(\sqrt{\Pi_i}\otimes\mathbb{1})|\Psi_{in}\ra}{\norm{(\sqrt{\Pi_i}\otimes\mathbb{1})|\Psi_{in}\ra}_1}=\sqrt{\frac{2^{N+1}}{\tr\Pi_i}}(\sqrt{\Pi_i}\otimes\mathbb{1})|\Psi_{in}\ra,
\ee where $|\Psi_{in}\ra = \otimes_{i=0}^{N}|\Psi^-\ra_{A_iB_i}$ and
\be
\norm{\li(\sqrt{\Pi_i}\otimes\mathbb{1}\pr)\Psi_{in}}_1^2 = \la\Psi_{in}|\Pi_i\otimes\mathbb{1}|\Psi_{in}\ra=\frac{1}{2^{N+1}}\tr\Pi_i.
\ee

Similarly, 
\be\label{idstate}
|\Psi_{id}^i\ra = \sqrt{2^{N-1}}\li(\sqrt{\sigma^{(i)}}\otimes\mathbb{1}\pr)|\Psi_{in}\ra.
\ee

The average fidelity has the form:
\be	
F\li(\precc\pr) = \sum_{i=1}^{N} p_iF(\rho_{out}^i, \Psi_{id}^{i})+p_0 F_{fail}\ge Np_1F(\rho_{out}^1, \Psi_{id}^{1}), 
\ee
where without loss of generality we can assume that Alice obtains outcome $z=1$, and the teleported state lands in the first port. Therefore, $p_1 = \tr(\Pi_1\frac{\mathbb{1}_A}{2^{N+1}}) = \frac{1}{2^{N+1}}\tr\Pi_1$, so 
\begin{align}\label{totalfidelity}
F\li(\precc\pr)&\ge \frac{N}{2^{N+1}}\tr(\Pi_1)F(\rho_{out}^1, \Psi_{id}^{1})\\
&=  \frac{N}{2^{N+1}}\tr(\Pi_1)\tr\li(\sqrt{\sigma^{(1)}_A}\sqrt{\frac{\Pi_1}{\tr \Pi_1}}\pr),
\end{align}

We further make some simplifications: $\sigma^{(1)}_A = \frac{1}{2^{N-1}}P^-$, where $P^-=\Psi^-_{A_0A_1}\otimes\mathbb{1}_{A_2...A_N}$. Thus 
\be
\sqrt{\sigma^{(1)}_A} = \sqrt{\frac{1}{2^{N-1}}}P^- =  \frac{2^{N-1}}{\sqrt{2^{N-1}}}\sigma^{(1)}_A = \sqrt{2^{N-1}}\sigma^{(1)}_A.
\ee
Therefore, $\tr\li(\sqrt{\sigma^{(1)}_A}\sqrt{\Pi_1}\pr) = \sqrt{2^{N-1}}\tr(\sigma^{(1)}_A\sqrt{\Pi_1})$. Substituting it in Eqn.~\eqref{totalfidelity} we get the resulting expression for fidelity:
\begin{align}
F\li(\precc \pr)\ge\frac{1}{4}\frac{N\sqrt{\tr\Pi_1}}{\sqrt{2^{N-1}}}\tr(\sigma^{(1)}_A\sqrt{\Pi_1}).
\end{align}

{\bf Proof of Theorem 1:}

We essentially need to compute $\tr\Pi_1$ and $\tr(\sigma^{(1)}_A\sqrt{\Pi_1})$. In particular, we show that $\tr\Pi_1\sim\frac{2^{N+1}}{N}$ and $\tr(\sigma^{(1)}_A\sqrt{\Pi_1})\sim\frac{2}{\sqrt{N}}$ up to the leading order. We provide detailed calculation for $\tr\Pi_1$, as $\tr(\sigma^{(1)}_A\sqrt{\Pi_1})$ can be obtained in a completely analogous way. We proceed by first representing the operators in the Schur basis, which simplifies the computation of the product of operators in each case. Then we explicitly compute the expressions to which both of them converge in distribution.

Consider the representation of the states $\rho_{out}$ and $\sigma^{(i)}_A$ in Schur basis of ${\cal H}^{\otimes N+1}$. In this representation we have a direct sum $\bigoplus_s{\cal H}_s$ of blocks with total spin $s$, each $\hs_s=\hs_{\da\da}\oplus\hs_{\ua\ua}\oplus\hs_{\ua\da}\oplus\hs_{\da\ua}$. The representation of $\rho_{out}, \sigma^{(1)}_A$ in this picture is
\begin{align}
\sigma^{(i)}_A &= \frac{1}{2^{N-1}}|\Psi^-_{A_0A_1}\ra\la\Psi^-_{A_0A_1}|\otimes\mathbb{1}_{A_2...A_N}\\
\rho_{out} &= \bigoplus_s\li[\lambda^+_s\li(Q_{\da\da}\oplus Q_{\ua\da}\pr)+\lambda_s^-\li(Q_{\ua\da}\oplus Q_{\da\ua}\pr)\pr],
\end{align}
where
\begin{align}
&Q_{\ua\ua}=\sum_{m,\beta}|\Psi_{\Rmnum 1}(\lambda_{s+\half}^+,m,\beta)\ra\la\Psi_{\Rmnum 1}(\lambda_{s+\half}^+,m,\beta)|,\\
&Q_{\ua\da}=\sum_{m,\beta}|\Psi_{\Rmnum 1}(\lambda_{s-\half}^-,m,\beta)\ra\la\Psi_{\Rmnum 1}(\lambda_{s-\half}^-,m,\beta)|,\\
&Q_{\da\ua}=\sum_{m,\beta}|\Psi_{\Rmnum 2}(\lambda_{s+\half}^+,m,\beta)\ra\la\Psi_{\Rmnum 2}(\lambda_{s+\half}^+,m,\beta)|,\\
&Q_{\da\da}=\sum_{m,\beta}|\Psi_{\Rmnum 2}(\lambda_{s-\half}^-,m,\beta)\ra\la\Psi_{\Rmnum 2}(\lambda_{s-\half}^-,m,\beta)|,
\end{align}
with $|\Psi_{\Rmnum 1 (\Rmnum{2})}(\lambda_{s\pm\half}^\pm,m,\beta)\ra$ defined in Eqn. (14-15) of~\cite{ishizaka_quantum_2009}. The parameter $\beta$ denotes the additional degree of freedom of the spin eigenbasis. As in the original protocol, we will consider only the space for irreps of the permutation group. 
From equations (19)-(23) in~\cite{ishizaka_quantum_2009} we get the following relations:
\begin{align}\label{exprforQst}
&Q_{\da\da}|\psi^-_{A_0A_1}\ra = 0,\\ 
&Q_{\ua\da}|\psi^-_{A_0A_1}\ra = \sqrt{\frac{s}{2s+1}}|\psi^-_{A_0A_1}\ra\otimes\mathbb{1}_{A_2...A_N},\\
&Q_{\da\ua}|\psi^-_{A_0A_1}\ra = -\sqrt{\frac{s+1}{2s+1}}|\psi^-_{A_0A_1}\ra\otimes\mathbb{1}_{A_2...A_N},\\
&Q_{\ua\ua}|\psi^-_{A_0A_1}\ra=0,\label{exprforQend}
\end{align}
where $\mathbb{1}_{A_2...A_n} = \bigoplus_s\mathbb{1}^s_{A_2...A_N}$. We will further identify $Q_{\ua\da} = Q_-$, and $Q_{\da\ua} = Q_+$. This leads to a more concise representation:
\begin{align}
\rho_{out} &= \bigoplus_s\mathbb{1}_U^s\otimes(\lambda_+Q_+\oplus\lambda_-Q_-)\\
\sigma^{(i)}_A &= \bigoplus_s \mathbb{1}_U^s\otimes R_i^s,
\end{align}
where $Q_\pm$ correspond to eigenspaces with eigenvalues $\lambda_{s\pm\half}^\pm$, and $R_i=\sum_\beta|\psi_\beta\ra\la\psi_\beta|$ is a projector such that $\la\psi_\beta|Q_+|\psi_\beta\ra = \frac{s}{2s+1}$, and $\la\psi_\beta|Q_-|\psi_\beta\ra = \frac{s+1}{2s+1}$. 
\\
To compute $\tr\Pi_1$ we first consider a more general expression: $X\widetilde\sigma^{(1)} X$, where $\widetilde\sigma^{(1)}_A = 2^{N-1}\sigma^{(1)}_A$, and operator $X$ can be written as:
\be
X =  \bigoplus_s\li[\gamma^+_sQ_+\oplus\gamma_s^-Q_-\pr].
\ee
From Eqn.~\eqref{exprforQst}-\eqref{exprforQend} it follows that:
\begin{align}\label{ysigmay}
X\widetilde\sigma^{(1)}&X =\\ 
&\bigoplus_{s,s'}[\gamma_s^+\gamma_{s'}^+Q^s_+\widetilde\sigma^{(1)}_AQ^{s'}_++ \gamma_s^+\widetilde\gamma_{s'}^-Q^s_+\widetilde\sigma^{(1)}_AQ^{s'}_-\\\nonumber
&+\gamma_s^-\gamma_{s'}^+Q^s_-\widetilde\sigma^{(1)}_AQ^{s'}_+ + \gamma_s^-\gamma_{s'}^-Q^s_-\widetilde\sigma^{(1)}_AQ^{s'}_-],
\end{align}
where we put superscripts $s, s{'}$ on top for clarity.
Note that by virtue of belonging to different irreps indexed by $s$, blocks with different eigenvalues are orthogonal, therefore all the terms in the direct sum where $s\neq s'$ will be zero. Evaluating each term individually, we get
\begin{align*}
&Q^s_+\widetilde\sigma^{(1)}_AQ_+^s = \frac{s}{2s+1}\sum_{m,\beta}|\Psi_{\Rmnum 1}^-\ra\la\Psi_{\Rmnum 1}^-|,\\
&Q^s_+\widetilde\sigma^{(1)}_AQ_-^s = -\frac{\sqrt{s(s+1)}}{2s+1}\sum_{m,\beta}|\Psi_{\Rmnum 1}^-\ra\la\Psi_{\Rmnum 2}^+|,\\
&Q^s_-\widetilde\sigma^{(1)}_AQ_+^s = -\frac{\sqrt{s(s+1)}}{2s+1}\sum_{m,\beta}|\Psi_{\Rmnum 2}^+\ra\la\Psi_{\Rmnum 1}^-|,\\
&Q^s_-\widetilde\sigma^{(1)}_AQ_-^s = \frac{s}{2s+1}\sum_{m,\beta}|\Psi_{\Rmnum 2}^+\ra\la\Psi_{\Rmnum 2}^+|,
\end{align*}
where 
\begin{align}
&|\Psi_{\Rmnum 1}^\pm\ra = |\Psi_{\Rmnum 1}(\lambda_{s\pm\half}^\pm,m,\beta)\ra,\\
&|\Psi_{\Rmnum 2}^\pm\ra = |\Psi_{\Rmnum 2}(\lambda_{s\pm\half}^\pm,m,\beta)\ra.
\end{align}
Only the terms where $Q$ has the same sign on the left and on the right from $\widetilde\sigma^{(1)}_A$ will result in non-zero contribution to the trace of Eqn.~\eqref{ysigmay}:
\begin{align}
\tr \li(X\widetilde\sigma^{(1)}_AX\pr) = &\tr(|\psi^-_{A_0A_1}\ra\la\psi^-_{A_0A_1}|\otimes\\
&\bigoplus_s[\li(\gamma_s^+\pr)^2\frac{s}{2s+1}+\li(\gamma_s^-\pr)^2 \frac{s+1}{2s+1}]\mathbb{1}^s_{A_2...A_N}).
\end{align}

 Recalling that $\gamma_s^\pm = \sqrt{\frac{1}{\lambda_s^\pm}}$, the explicit form for $\tr\Pi_1$ is:
 \be\label{someeq}
 \tr\Pi_1 =\frac{1}{2^{N-1}}\sum_{s=s_m}^{\frac{N-1}{2}} \li(\frac{1}{\lambda_s^-}\frac{s}{2s+1}+\frac{1}{\lambda_s^+}\frac{s+1}{2s+1}\pr) d_s(N-1),
 \ee
 where $s_m=0(\half)$  when $N$ is odd(even), and $d_s(N-1) = \mbox{dim}(\hs_U^s\otimes\hs_P^s) = (2s+1)g_s(N-1)$. From the equations (7), (9) of~\cite{ishizaka_quantum_2009} we know that 
 \begin{align*}
 &g_s(N-1) = \frac{(2s+1)(N-1)!}{(\frac{N-1}{2}-s)!(\frac{N-1}{2}+1+s)!},\\
 &\lambda_s^+\equiv\lambda^+_{s+\half}=\frac{1}{2^{N}}\li(\frac{N}{2}+s+\frac{3}{2}\pr),\\
 &\lambda_s^-\equiv\lambda^-_{s-\half}=\frac{1}{2^{N}}\li(\frac{N}{2}-s+\half\pr).
 \end{align*}
 Finally, substituting them in the Eqn.~\eqref{someeq}:
 \begin{widetext}
\be
\tr\Pi_1 = \tr\rho^{-\half}\sigma^{(1)}\rho^{-\half} = 2\sum_{s=s_m}^{\frac{N-1}{2}}\li(\frac{s}{\frac{N}{2}-s+\frac{1}{2}}+\frac{s+1}{\frac{N}{2}+s+\frac{3}{2}}\pr)\frac{(2s+1)(N-1)!}{(\frac{N-1}{2}-s)!(\frac{N-1}{2}+s+1)!},
\ee
 \end{widetext}
We will further compute $ \tr\Pi_1$ up to terms $O\li(\frac{1}{N^2}\pr)$ by first symmetrically extending the corresponding sums to the range $s\in[0,N]$ for odd $N$ and $s\in[\half,N]$ for even. Then we note that the obtained expressions are the expectation values of a function of the random variable with binomial distribution. Then, using de Moivre--Laplace theorem~\cite{feller_introduction_1968} in the limit of large $N$ we replace the binomial random variables by the ones distributed as $N(0,1)$. Then, we compute the expectations directly, obtaining the result:
\begin{widetext}
\begin{align}
\tr\Pi_1 &= 2\sum_{s=0}^{\frac{N-1}{2}}\li(\frac{s}{\frac{N}{2}-s+\frac{1}{2}}+\frac{s+1}{\frac{N}{2}+s+\frac{3}{2}}\pr)\frac{(2s+1)(N-1)!}{(\frac{N-1}{2}-s)!(\frac{N-1}{2}+s+1)!}\label{firstlineofcalc}\\
&=\frac{2^{N-1}}{N}\sum_{k=0}^{N}\li(\frac{N-2k-1}{k+1}+\frac{N-2k+1}{N-k+1}\pr)(N-2k){N \choose k}\frac{1}{2^N}\label{secondline}\\
&\approx\frac{2^{N-1}}{N}\mathbb{E}\li[\li(\frac{\sqrt{N}Z+1}{\frac{N}{2}+\frac{\sqrt{N}}{2}Z+1}+\frac{\sqrt{N}Z-1}{\frac{N}{2}-\frac{\sqrt{N}}{2}Z+1}\pr)\sqrt{N}Z\pr]\label{thirdline}\\
&=\frac{2^{N}}{N^2}\mathbb{E}\li[\li( (\sqrt{N}Z+1)\li(\frac{Z}{\sqrt{N}}+\frac{2}{N}+1\pr)^{-1}+(\sqrt{N}Z-1)\li(-\frac{Z}{\sqrt{N}}+\frac{2}{N}+1\pr)^{-1}\pr)\sqrt{N}Z\pr]\label{fourthline}\\
&=\frac{2^{N}}{N^2}\mathbb{E}\li[\frac{Z}{N^{\frac{3}{2}}}\li((N(N-2)-4)+ \sqrt{N}(N-4)(2N+3)Z-NZ^2-2N^{\frac{3}{2}}Z^3+O\li(\frac{1}{\sqrt{N}} \pr)\pr)\pr]\label{secondorder}\\
&=\frac{2^{N}}{N^2}\frac{1}{N^{\frac{3}{2}}}\li(\sqrt{N}(N-4)(2N+3)-6N^{\frac{3}{2}} + O\li(\frac{1}{\sqrt{N}}\pr)\pr)\label{penultline}\\
&=\frac{2^{N+1}}{N}\li(1-\frac{11}{2N}-\frac{6}{N^2}+O\li(\frac{1}{N^3}\pr)\pr),
\end{align}
\end{widetext}

In~\eqref{secondline} we used the substitution $s = \frac{N}{2}-(k+\half)$ and symmetrized the sum to include the range $[\frac{N+1}{2},N]$, moving the discrete probability function to the right side of the sum.
We note that Eqn.~\eqref{thirdline} is the expectation of the function
\be
f_N(k):=\li(\frac{N-2k-1}{k+1}+\frac{N-2k+1}{N-k+1}\pr)(N-2k)\nonumber
\ee
of the random variable $k\sim\text{Binom}(N,\half)$.
In the limit of large $N$, $k$ is well approximated by the random variable $Z\sim N(0,1)$:
\be
k\approx_d \frac{N}{2}+\frac{\sqrt{N}}{2}Z.
\ee
Then, in Eqns.~\eqref{fourthline},~\eqref{secondorder} we simplify the expression inside the expectation using the expansion $(1+x)^{-1}\approx 1- x + x^2$, which is valid when $x\to 0$ (this holds when $N\to\infty$). Lastly, in Eqn.~\eqref{penultline} we use the fact that $\mathbb{E}[Z^{2m+1}]=0$, and $\mathbb{E}[Z^{2m}]=(2m-1)!!$.

We now turn to compute $\tr\li(\sigma^{(1)}_A\sqrt{\Pi_1}\pr)$:
\begin{align}
&\tr\li(\sigma^{(1)}_A\sqrt{\Pi_1}\pr)\\&=\tr\li(\sigma^{(1)}_A\sqrt{\rho^{-\half}\sigma^{(1)}_A\rho^{-\half}}\pr)\\
&=\sqrt{2^{N-1}}\tr\sum_{s=s_m}^{\frac{N-1}{2}}\sigma^{(1)}_A(s)\rho^{-\frac{1}{4}}(s)\sigma^{(1)}_A(s)\rho^{-\frac{1}{4}}(s)\label{purestsimp}\\
&=\frac{1}{2^{N-1}}\frac{1}{\sqrt{2^{N-1}}}\sum_{s=s_m}^{\frac{N-1}{2}}c(s,4)(2s+1)g_s(N-1),
\end{align}
where $c(s,y) = \frac{s}{2s+1}\li(\lambda_{s-\half}^{-}\pr)^{-\frac{1}{y}}+ \frac{s+1}{2s+1}\li(\lambda_{s+\half}^{+}\pr)^{-\frac{1}{y}}$. In Eqn.~\eqref{purestsimp} we make use of the equality $\sqrt{\sigma^{(1)}_A}=\sqrt{2^{n-1}}\sigma^{(1)}_A$. Substituting values of $\lambda^\pm_{s\pm\half}$ and simplifying the calculation analogously to~\eqref{firstlineofcalc}-~\eqref{penultline} we get:
\begin{widetext}
\begin{align}
\tr\li(\sigma^{(1)}_A\sqrt{\Pi_1}\pr)&=\\&=2\sqrt{2}\sum_{s=s_m}^{\frac{N-1}{2}}\li(\li(\frac{N}{2}+s+\frac{3}{2}\pr)^{-\frac{1}{4}}(s+1)+\li(\frac{N}{2}-s+\frac{1}{2}\pr)^{-\frac{1}{4}}s\pr)^2\frac{(N-1)!}{\li(\frac{N-1}{2}-s\pr)!\li(\frac{N-1}{2}+s+1\pr)!}\frac{1}{2^{N-1}}\\
&=\frac{\sqrt{2}}{4N}\sum_{k=0}^{N}\li(\frac{N-2k+1}{(N-k+1)^{-\frac{1}{4}}} + \frac{N-2k-1}{(k+1)^{-\frac{1}{4}}}\pr)^2 {N\choose k}\frac{1}{2^N}\\
&=\frac{\sqrt{2}}{4N}\mathbb{E}\li(\frac{N-2k+1}{(N-k+1)^{\frac{1}{4}}} + \frac{N-2k-1}{(k+1)^{\frac{1}{4}}}\pr)^2\\
&\approx_d \frac{\sqrt{2}}{4N}\mathbb{E}\li(\frac{\sqrt{N}Z-1}{\li(\frac{N}{2}-\frac{\sqrt{N}}{2}Z+1\pr)^{\frac{1}{4}}} + \frac{\sqrt{N}Z+1}{\li(\frac{N}{2}+\frac{\sqrt{N}}{2}Z+1\pr)^{\frac{1}{4}}}\pr)^2\\
&=\frac{1}{2N^{\frac{3}{2}}}\mathbb{E}\li(\li(\sqrt{N}Z-1\pr)\li(1-\frac{Z}{\sqrt{N}}+\frac{2}{N}\pr)^{-\frac{1}{4}} 
								\li(\sqrt{N}Z+1\pr)\li(1+\frac{Z}{\sqrt{N}}+\frac{2}{N}\pr)^{-\frac{1}{4}}\pr)^2\\
&=\frac{1}{2N^{\frac{3}{2}}}\li(4N+\frac{39}{4}+O\li(\frac{1}{N}\pr)\pr)\\
&=\frac{2}{\sqrt{N}} + O\li(\frac{1}{N^{3/2}}\pr)
\end{align}
\end{widetext}

Finally, 
\begin{align}
F\li(\precc (N,k)\pr)&= \frac{N}{4}\sqrt{\frac{\tr\Pi_1}{2^{N-1}}}\tr\li(\sigma^{(1)}_A\sqrt{\Pi_1}\pr)\\
&=1-\frac{11}{4N}+O\li(\frac{1}{N^2}\pr).
\end{align}
\qed

\subsection{Lemma 2}
In what follows without loss of generality we will assume that the teleported state goes to port 1. We will also use the notation: $\rho_{out,AB}=\rho^1_{out}$ and $\Psi_{id,AB}= \Psi^1_{id}$.
From Theorem 1 it follows that the fidelity of the overall state after the teleportation 
\be\label{lowerbound}
F\li(\rho_{out,AB}, \Psi_{id,AB}\pr)\ge1-\frac{11}{4N},
\ee
where ${AB}=A_0B_0A_1B_1...A_NB_N$, with $\Psi_{id,AB}$ defined in Eqn.~\eqref{idstate}. Unlike the original teleportation protocol~\cite{ishizaka_quantum_2009} where the subsystems $A_2B_2...A_NB_N$ were traced out, we keep them to use in the following rounds. However, it is necessary to understand how much noise is picked up by each port individually to rule the situation where the state of each port degrades disproportionately to others, affecting the state, which lands in that port. Using the fact that fidelity does not increase under partial trace, it turns out that the state of the ports are close to $N-1$ copies of EPR state:
\begin{align}\label{porterrorbound}
1-\frac{11}{4N}&\le F\li(\rho_{out,AB},\Psi_{id,AB}\pr)\\
&\le F\li(\rho_{{out,AB}\backslash\{{A_0B_0}\}},\Psi_{{id,AB}\backslash\{{A_0B_0}\}}\pr).
\end{align}
Therefore, it follows that the fidelity of each individual port with singlet after the teleportation has the same lower bound as~\eqref{lowerbound}.

{\bf Proof of Lemma 2:}

May the total state at the beginning of the protocol be 
\be
\Omega_{0}=\omega_{C_1...C_k}\otimes\phi_{A_1B_1...A_NB_N},
\ee
where $\omega_{C_1...C_k}$ denote a $k$-qubit state that Alice wants to teleport to Bob using $\precc$, and $\phi_{A_1B_1...A_NB_N}$ denote the ports. After the first round of the teleportation protocol the state of the qubit in subsystem ${C_1}$ goes to $A_1$, and the total state becomes:
\be
\Omega_1 = \theta_{C_1...C_kA_1B_1...A_NB_N}.
\ee
From Theorem 1 we know that
\be\label{recconseq}
\norm{\theta_{A_2B_2...A_NB_N}-\li(\Psi^-\pr)^{\otimes N-1}}_1\le\frac{11}{4N}.
\ee
After teleportation, the subsystems $A_2B_2...A_NB_N$ may be entangled with the teleported state. In order to decorrelate them, Alice and Bob apply the following operation:
\be
\Delta(\rho) = \int_{U(d)} dU\li(U\otimes U\pr)\rho\li(U\otimes U\pr)^\dagger
\ee

 to the subsystems $A_2B_2...A_NB_N$. This results in
 \begin{align*}
 \Omega_{tw} &= \li(\id_{C_1...C_k}\otimes\Delta\pr)(\Omega_1)\\
 &=(1-p)\pi_{C_1...C_k}\otimes \phi_{A_2B_2...A_NB_N} \\&+ p\sigma_{C_1...C_k}\otimes \ID_{A_2B_2...A_NB_N},
 \end{align*}
 where $p\le\frac{11}{4N}$. From the port-based teleportation protocol we know that
 \be
 \norm{\theta_{C_1...C_k}-\omega_{C_1...C_k}}_1\le \frac{11}{4N},
 \ee
because we teleport only the $C_1$ subsystem, and leave other intact, hence 
 \be
 \norm{(1-p)\pi_{C_1...C_k} + p\sigma_{C_1...C_k}-\omega_{C_1...C_k}}_1\le \frac{11}{4N},
 \ee
 which, combined with~\eqref{recconseq} gives
 \be
 \norm{\pi_{C_1...C_k}-\omega_{C_1...C_k}}_1\le \frac{11}{2N}.
 \ee
 Repeating the teleportation $k$ times we obtain the statement of the Lemma.
 \qed
\section{3. Simultaneous teleportation}
To get the lower bounds on the efficiency of this protocol we use the observation that the task of teleporting an unknown quantum state and signal discrimination are equivalent, as shown in~\cite{bk2011} (see also~\cite{damian} for a different treatment of port-based teleportation). There, authors show that the average fidelity of teleportation corresponds to the average fidelity of discriminating a set of signals with each signal representing different ports where teleported state arrives to. Given that we want to teleport multiple states at once, our set of states $\{\eta^g_k \}_{g\in S_N}$ corresponds to all possible ways in which $k$ teleported systems arrive in $N$ ports:
\be
\eta^g_k = U_g\li(\tr_{B_1...B_N\backslash B_{g(1)}...B_{g(k)}}\li[\bigotimes_{i=1}^N|\Phi^+\ra\la\Phi^+|_{A_iB_i}\pr]\pr)U_g^\dagger,
\ee
where $U_g$ permutes the subsystems according to some $g\in S_N$ -- symmetric permutation group on $N$ elements.

{\bf Proof of Theorem 2:}
we prove the result of the Theorem for qudits, where Alice wants to teleport the subsystems $A_1....A_k$ of the state $\sigma_{A_1B_1...A_kB_k}$. To simplify the calculations we take $\sigma = \otimes_{j=0}^{k}|\Psi^-_{A_jB_j}\ra$.
The total number of POVM elements that Alice has at her disposal is $M=k!{N\choose k}=\frac{N!}{(N-k)!}$.
We now compute the fidelity of successful teleportation by estimating

\be
F(\psimfull N k)=\frac{M}{d^{2k}}p_s,
\ee

where $p_s\ge\frac{1}{M\bar{r}\tr \bar\eta^2_k}$. From the structure of states $\eta^g_k$ it follows that $\bar r = \frac{1}{M}\sum_{g\in S_N}\text{rank}(\eta^g_k)=\eta^{g_0}_k=\frac{1}{d^{N-k}}$, as all states $\eta^g_k$ have the same rank. What remains to be computed is $\tr\bar\eta^2_k=\frac{1}{M}\sum_{g\le g'}\tr\li(\eta^g_k\eta^{g'}_k\pr)$. 
To compute $\tr\li(\eta^g_k\eta^{g'}_k\pr)$ we need to look on a few different cases depending on the relative location of maximally entangled states in the corresponding states.
We use the following notation: $'-'$ denotes the $\ID/d$ subsystem in the tensor product, $|xx|$ denotes $\Phi^+$. 

Consider a particular case: fix $\eta^g_k = ---|xx|xx|--$ and vary $\eta^{g'}_k$ ($N=7, k=2$). The following table illustrates the possible values of the trace of the pairwise product, depending on the relative position of the maximally entangled states in $\eta^g_k$, with respect to $\eta^{g'}_k$. The results are easily generalized for the case of arbitrary $k$.
\begin{table}[h]
\centering
\caption{The effect of relative position of the MES on the value of trace.}
\label{tableofoverlaps}
\begin{center}
    \begin{tabular}{ | c | c |}
    \hline
    $\eta^{g'}_k$ & $\tr\li(\eta^g_k\eta^{g'}_k\pr)$ \\ \hline
    $|xx|x{\bf x}|-----$ & $\frac{1}{d^{N+k}}$ \\ \hline
    $-|xx|{\bf xx}|----$ & $\frac{1}{d^{N+k-2}}$ \\ \hline
     $--|x{\bf x|xx}|---$ & $\frac{1}{d^{N+k}}$ \\ \hline
     $---|{\bf xx|xx}|--$ & $\frac{1}{d^{N+k-4}}$\\ \hline
    \end{tabular}
\end{center}
\end{table}
Subsystems in bold denote partial (odd number of subsystems) or full (even number of subsystems) overlaps between $\eta^g_k$ and $\eta^{g'}_k$. The cases with partial overlaps contribute $\frac{1}{d^{N+k}}$ -- in the same way as no overlap at all. And in case of full overlaps, we subtract from the exponent twice the number $t$ of overlapping MES, yielding $\frac{1}{d^{N+k-2t}}$.
The expression $\tr\bar\eta^2_k$ consists of the sum of the terms $\frac{1}{d^{N+k-2t}}$ for some $t\in\{0,...,k\}$, depending on the index of the permutation. We will  determine the multiplicity of each of the terms indexed by $t$. For this, fix $g$, and consider all $g'\neq g$ that result in 
\be\label{singleT}
\tr\li(\eta^g_k\eta^{g'}_k\pr)=\frac{1}{d^{N+k-2t}}
\ee
for some fixed $t$. There are $t! {k\choose t}$ ways to pick $g'$ such that the there are $t$ full overlaps. In order to position the remaining $k-t$ maximally entangled states in $\eta^{g'}_k$, we must take into the acount $k-t$ 'forbidden' positions of $\eta^g_k$ (the location of the remaining maximally entangled states), where there must be no full overlaps, but only partial ones. This leaves $N+k-2t-(k-t)=N-t$ 'good' positions. Accounting for all possible permutations of the remaining maximally entangled states, there are $(k-t)!{{N-t}\choose{k-t}}$ ways to fill the rest of the good positions. Therefore, the total number of $g'$ that for fixed $g, t$  results in~\eqref{singleT} is 
\be
L_{N,k,t}=t! {k\choose t}(k-t)!{{N-t}\choose{k-t}}=\frac{k!(N-t)!}{(k-t)!(N-k)!}.
\ee
Due to symmetry, $L_{N,k,t}$ is the same for each $g$.
Finally, 
\begin{align}
\tr\bar\eta^2_k&=\frac{1}{M^2}\li(\sum_g\tr\eta^g_k\eta^g_k + \sum_{g\neq g'}\tr\eta^g_k\eta^{g'}_k\pr)\\
&=\frac{1}{Md^{N-k}}+\frac{1}{M}\sum_{t=1}^{k} \frac{k!(N-t)!}{(k-t)!(N-k)!}\frac{1}{d^{N+k-2t}}\label{GP}\\
&=\frac{1}{Md^{N-k}}+\frac{1}{M}\frac{k!N!}{(N-k)!k!}\li(\frac{1-\li(\frac{k}{N}d^2\pr)^k}{1-\frac{k}{N}d^2}\pr)\frac{1}{d^{N+k}}\\
&=\frac{1}{Md^{N-k}}+\li(\frac{1-\li(\frac{k}{N}d^2\pr)^k}{1-\frac{k}{N}d^2}\pr)\frac{1}{d^{N+k}},
\end{align}

where we used $M=\frac{N!}{(N-k)!}$, and noted that~\eqref{GP} is geometric progression with $b_0=\frac{N!}{k!}\frac{1}{d^{N+k}}, r=\frac{k}{N}d^2$. The resulting fidelity of teleportation has the form:
\begin{widetext}
\begin{align}
F(\psimf)&\ge \frac{M}{d^{2k}}\frac{1}{Md^{N-k}}\li[\frac{d^{2k}}{M}+\li(\frac{1-\li(\frac{k}{N}d^2\pr)^k}{1-\frac{k}{N}d^2}\pr)\pr]^{-1}d^{N+k}\\
&=\li[\frac{d^{2k}}{M}+\li(\frac{1-\li(\frac{k}{N}d^2\pr)^k}{1-\frac{k}{N}d^2}\pr)\pr]^{-1}.
\end{align}
\end{widetext}
\cite{bennett_remote_2005}
Setting $d=2, k=\epsilon N$ and taking limit $N\to\infty$ we obtain the result.
\qed

\bibliographystyle{abbrv}

\end{document}